\documentstyle[11pt,fleqn,epsf]{article}
\iftrue \textheight47\baselineskip 
\else\textheight22.47cm \fi
\begin{document}
\sloppy
\title{\begin{flushright}\normalsize HD-THEP-94-38\end{flushright}
\Huge \bf SU(3) Flux Tubes in a Model of the stochastic Vacuum}
\author{
Michael Rueter\thanks{e-mail: rueter@next3.thphys.uni-heidelberg.de}\\[.5cm]
\it Institut f\"ur theoretische Physik\\
\it Universit\"at Heidelberg\\
\it Philosophenweg 16, D-69120 Heidelberg, FRG\\[1cm]
\and H.G. Dosch\\[.5cm]
\it Institut f\"ur theoretische Physik\\
\it Universit\"at Heidelberg\\
\it Philosophenweg 16, D-69120 Heidelberg, FRG
}
\date{}
\maketitle
\thispagestyle{empty}
\begin{abstract}
We calculate the squared gluon field strengths of a heavy q-$\rm \bar{q}$-pair
in the model of the stochastic vacuum. We observe that with increasing
separation a chromoelectric flux tube is built. The properties of the emerging
flux tube are investigated.
\end{abstract}
\newpage
\setcounter{page}{1}
\section{Introduction}
As is well known, the behavior of QCD at large distances is most probably not
controlled by perturbation theory, but genuine non-perturbative effects are
supposed to play an essential role in that infrared domain. This is indicated
as well by analytical considerations \cite{3} as by numerical calculations from
lattice-QCD \cite{4}. It is of course the ultimate desire to obtain analytical
results directly from the QCD-Lagrangian, but successes in that direction are
limited. At the moment all analytical results are based on models.\\
In the model of the stochastic vacuum (MSV) it is assumed that the complicated
measure which emerges from the QCD-action in the region of slowly fluctuating
fields can be described by a well behaved stochastic process \cite{24} \cite{1}
\cite{21}. This process can be visualized as a fluctuating background field
representing the "physical vacuum".\\
In the most general form of the MSV one assumes that the above mentioned
background field has a convergent cluster expansion, but for many specific
applications phenomenologically useful results can only be obtained after
further assumptions are made. One of these assumptions is that the stochastic
vacuum can be described by a Gaussian process, i.e.~only the 2-cumulant does
not vanish. In a non-Abelian theory even the assumption of a Gaussian process
has to be specified further, since the choice of the stochastic variables is
ambiguous.

Our paper is organized as follows:\\
In section 2 we shortly present the model of the stochastic vacuum in a
restricted form as it has been applied very successfully to soft hadron-hadron
high-energy scattering \cite{29} \cite{59}.\\
In section 3 we apply this restricted form of the model in order to calculate
the squares of the field strengths of Euclidean Wegner-Wilson-loops. Especially
we shall test the importance of some more technical assumptions described in
section 2 and check the consistency of the approximations by investigating the
Michael sum rule of the total energy stored in the field of the
Wegner-Wilson-loop.\\
In section 4 we shall present the results, discuss them and compare, where
possible, with lattice gauge calculations.

Throughout the paper we work in an {\bf Euclidean} space-time continuum.
\section{The model of the stochastic vacuum}
A rather extensive description of the MSV as applied in this paper has been
given in a recent publication \cite{59}, so we quote here for convenience only
the main points.\\
The crucial ingredient of the MSV is the correlator between the gluon field
strengths $F_{\mu\nu}^A(x,w)$ parallel transported to a reference point $w$
\begin{equation}
\label{tran Feld}
{\bf F}_{\mu \nu}(x,w) \; = \; {\bf \Phi}^{-1}(x,w) \: {\bf F}_{\mu \nu}(x) \:
{\bf \Phi}(x,w),
\end{equation}
where ${\bf F}_{\mu\nu}(x)$ is the Lie-algebra valued gluon field strength
tensor and ${\bf \Phi}(x,w)$ the path ordered non-Abelian Schwinger string:
\begin{eqnarray}
{\bf \Phi}(x,w) \; &=& \; {\cal P} \exp \left[ -ig \int_0^1 d\sigma \:
(x-w)_\mu {\bf A}_\mu(w+\sigma (x-w)) \right]\nonumber \\
{\bf F}_{\mu\nu}(x)\; &=& \; \sum_{C=1}^{N_c^2-1}F_{\mu\nu}^C(x)\: {\bf t}^C
\end{eqnarray}
$N_c$ is the number of colors and unless otherwise stated we use the
fundamental representation.\\
We make the crucial approximation that the correlator $<F_{\mu\nu}^A(x,w)\:
F_{\rho\sigma}^B(y,w)>_A$, i.e.~the vacuum expectation value with respect to
the non-perturbative background field, is independent of the reference point
$w$. Then the most general form of the correlator is:
\pagebreak[0]
{\mathindent0em
\begin{eqnarray}
\label{Korrelator}
<g^2 F_{\mu \nu}^A (x,w) \: F_{\rho\sigma}^B(y,w)> \; = \; \frac{\delta
^{AB}}{N_c^2-1}\frac{<g^2 FF>}{12} \bigg\{ (\delta_{\mu\rho}
\delta_{\nu\sigma}\: - \: \delta_{\mu\sigma}\delta_{\nu\rho}) \kappa \:
D(z^2/a^2) \nonumber \\
+ \frac{1}{2}\left( \frac{\partial}{\partial z_\nu}(z_\sigma \delta_{\mu\rho} -
z_\rho \delta_{\mu\sigma})+\frac{\partial}{\partial z_\mu}(z_\rho
\delta_{\nu\sigma} - z_\sigma \delta_{\nu\rho}) \right) (1-\kappa )D_1(z^2/a^2)
\bigg\}
\end{eqnarray}}\parindent0em
where $z=x-y$.\\
The fundamental parameters of the model are the correlation length $a$ and the
gluon condensate $<g^2FF>\; = \; <g^2 F_{\mu\nu}^C\: F_{\mu\nu}^C>$.\\
The factors in eq.(\ref{Korrelator}) are chosen in such a way that $D(0)\; = \;
D_1(0)\; = \; 1.$\\
By evaluating the expectation value of a Wegner-Wilson-loop with the correlator
(\ref{Korrelator}) and Gaussian factorization one obtains the potential for a
static q-$\rm \bar{q}$-pair \cite{26}
\begin{eqnarray}
\label{Potential}
V(r) \; &=& \; \frac{<g^2FF>a^3}{48N_C} \bigg\{ 2r \int_0^r d\rho
\int_{-\infty}^\infty d\tau \;\kappa\; D(\rho^2 +\tau^2) \nonumber \\
&&+\int_0^r \rho\: d\rho \int_{-\infty}^\infty d\tau \left( -2\kappa D(\rho^2
+\tau^2)+(1-\kappa )D_1(\rho^2 +\tau^2)\right) \bigg\}.
\end{eqnarray}\parindent0em
In an Abelian gauge theory without monopoles, where the homogeneous Maxwell
equation must hold, only the second structure in eq.(\ref{Korrelator}) can
occur, i.e.~we must have $\kappa=0$. However, in a non-Abelian theory there is
no reason for $\kappa$ to be zero and it is the tensor structure proportional
to $\kappa$ which leads to the linear rising potential (first term in equation
\ref{Potential}) and hence confinement.\\
{}From comparison with hadron spectroscopy, high-energy scattering \cite{59}
and lattice gauge calculations \cite{33} the input to the right hand side of
eq.(\ref{Korrelator}) is pretty well fixed:
\begin{eqnarray}
D( z^2/a^2) \; &=& \; -6 A_4 \int \frac{d^4 k}{(2\pi)^4} \frac{k^2}{(k^2+1)^4}
\exp \left( i\: \frac{k\: (z/a)}{\lambda_4}\right) \nonumber\\
&=& \; \frac{|z/a|}{\lambda_4} \left[ K_1\left( \frac{|z/a|}{\lambda_4} \right)
- \frac{1}{4} \frac{|z/a|}{\lambda_4}K_0\left( \frac{|z/a|}{\lambda_4} \right)
\right] \nonumber \\
D_1( z^2/a^2) \; &=& \; -4 A_4 \int \frac{d^4 k}{(2\pi)^4} \frac{1}{(k^2+1)^3}
\exp \left( i\: \frac{k\: (z/a)}{\lambda_4}\right) \nonumber\\
&=& \; \frac{|z/a|}{\lambda_4} K_1\left( \frac{|z/a|}{\lambda_4} \right)
\end{eqnarray}
with $\lambda_4=\frac{8}{3\pi}$, $A_4=-8\pi^2$, $a=0.35$fm, $<g^2FF>= 2.39{\rm
GeV}^4$ and $\kappa=0.74$.\\
The $K_\mu$ are the modified Bessel-functions.

\epsfxsize8cm
\begin{figure}[ht]
\leavevmode
\centering
\epsffile{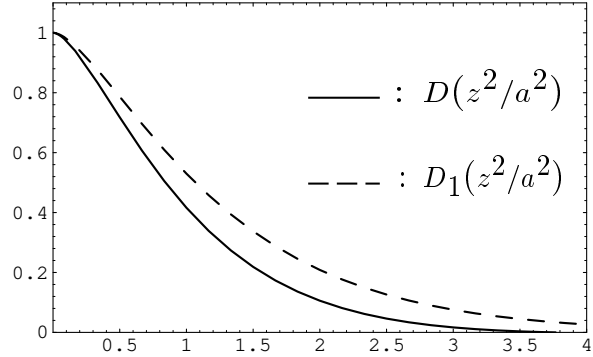}
\caption{The functions $D(z^2/a^2)$ and $D_1(z^2/a^2)$ of eq.(5)}
\end{figure}
\section{The squared components of the gluon field tensor}
\begin{figure}[ht]
\leavevmode
\centering
\epsffile{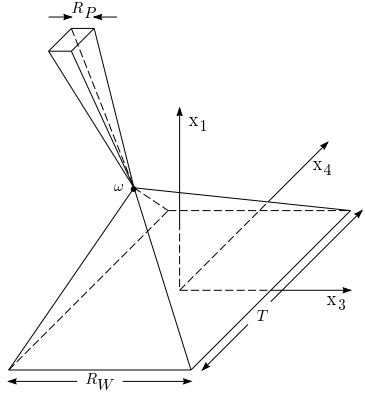}
\caption{The two pyramids touching each other at the reference point $w$}
\label{Pyramide}
\end{figure}

In this section we evaluate the tensor of the squared field strengths in the
presence of a Wegner-Wilson-loop.\\
We introduce as usual the Wegner-Wilson-loop over a contour $\cal C$
\begin{equation}
<W[{\cal C} ]>\; = \; tr\:{\cal P} < \exp \left\{ -ig \oint_{\cal C} \mbox{\bf
A}_\mu (x) dx_\mu \right\} >
\end{equation}
The plaquette $P_{\alpha\beta}(x)$ is a Wegner-Wilson-loop over a square in
$\alpha,\beta-$direction centered at $x$ with side length $R_P$.\\
One easily obtains
\begin{equation}
P_{\alpha\beta}(x)\; = \; N_c-\frac{1}{4} R_P^4\:g^2\: \sum_C
F_{\alpha\beta}^C(x)F_{\alpha\beta}^C(x)\; +\; {\cal O}(R_P^8)
\end{equation}
where there is {\bf no summation} over $\alpha$ and $\beta$ on the r.h.s.\\
The non-perturbative vacuum expectation value of $P_{\alpha\beta}$ is related
to the gluon condensate:
\begin{equation}
\sum_{\alpha<\beta}<P_{\alpha\beta}>\; = \; 6N_c-\frac{1}{8}R_P^4\: <g^2\:
FF>\;+\; {\cal O}(R_P^8)
\end{equation}
To calculate the square of the gluon field strengths in the presence of a
static quark-antiquark-pair we define the quantity \cite{38} \cite{39}
\begin{equation}
\label{Def F}
\Delta F^2_{\alpha\beta}\:(x)\; \equiv \; \frac{4}{R_P^4} \frac{<W[{\cal C}]\:
P_{\alpha\beta}(x)>-<W[{\cal C}]><P_{\alpha\beta}(x)>}{<W[{\cal C}]>}.
\end{equation}
Here $\cal C$ is the Wegner-Wilson-loop corresponding to a static q-$\rm
\bar{q}$-pair with side length $R_W$ in 3-direction and $T$ in 4-direction
centered at the origin (see figure \ref{Pyramide}).\\
In the limit $R_P \to 0$ the elements of the tensor (\ref{Def F}) are the
differences between the expectation values of the squared elements of the gluon
field strength tensor in the presence of the q-$\rm \bar{q}$-pair and the
fluctuations of the non-perturbative vacuum.\\
In order to evaluate $\Delta F^2_{\alpha\beta}\:(x)$ in the model of the
stochastic vacuum we first transform as usual the path ordered line-integrals
in the Wegner-Wilson-loops into surface ordered surface-integrals with the help
of the non-Abelian Stokes-theorem \cite{57}.
\begin{eqnarray}
<W[{\cal C} ]>\; &=& \; tr\: {\cal P} < \exp \left\{ -ig \oint_{\cal C}
\mbox{\bf A}_\mu (x) dx_\mu \right\} >\nonumber \\
&=& \; tr\: {\cal P}_{\cal S} < \exp \left\{ -ig \int_{\cal S} \mbox{\bf
F}_{\mu\nu} (x,w) d\sigma_{\mu\nu} \right\} >
\end{eqnarray}
The surface ${\cal S}$ is boardered by the loop ${\cal C}$ and contains the
reference point $w$. The indices of the surface elements are restricted to $\mu
< \nu$ throughout this paper.\\
Expanding the exponentials in eq.(\ref{Def F}) we obtain:
{\mathindent0em
\begin{eqnarray}
\label{Ber F 1}
\Delta F^2_{\alpha\beta}\:(x) &=& \frac{4}{R_P^4} \frac{1}{<W>} \Bigg(
\sum_{n=1}^\infty (-i)^n \stackrel{\rm surface ordered}{\int \cdots \int}
d\sigma_{\mu_1\nu_1}^W\cdots d\sigma_{\mu_n\nu_n}^W\: tr\left[ {\bf
t}^{a_1}\cdots {\bf t}^{a_n} \right] \times \nonumber \\
&& \int\int d\sigma_{\mu\nu}^P d\sigma_{\rho\sigma}^P \times
\frac{(-i)^2}{2!}tr \left[ {\bf t}^a {\bf t}^b \right] <g^n
F_{\mu_1\nu_1}^{a_1} \cdots F_{\mu_n\nu_n}^{a_n}\: g^2
F_{\mu\nu}^{a}F_{\rho\sigma}^{b}>\; - \nonumber \\
&&\sum_{n=1}^\infty (-i)^n \stackrel{\rm surface ordered}{\int \cdots \int}
d\sigma_{\mu_1\nu_1}^W\cdots d\sigma_{\mu_n\nu_n}^W\: tr\left[ {\bf
t}^{a_1}\cdots {\bf t}^{a_n} \right] \: \int\int d\sigma_{\mu\nu}^P
d\sigma_{\rho\sigma}^P \times \nonumber \\
&&\frac{(-i)^2}{2!}tr \left[ {\bf t}^a {\bf t}^b \right] <g^n
F_{\mu_1\nu_1}^{a_1} \cdots F_{\mu_n\nu_n}^{a_n}>\; <g^2
F_{\mu\nu}^{a}F_{\rho\sigma}^{b}> \Bigg)
\end{eqnarray}}\parindent0em
Here we have neglected the higher terms in the expansion of $P_{\alpha\beta}$
since they are in higher order of $R_P$. In eq.(\ref{Ber F 1}) we have two
loops and hence two surfaces each containing the same point $w$. We have thus
chosen the surfaces as sliding sides of a pyramid based either on the
Wegner-Wilson-loop ($d\sigma^W$) or on the plaquette ($d\sigma^P$) (see figure
\ref{Pyramide}).

\epsfxsize5cm
\parindent0em
We use the factorization of the color components of the field strength tensor
as in \cite{27} and \cite{28}, i.e. for example
{\mathindent0em\begin{eqnarray}
\label{Faktorisierung}
<F^{C_1}(1)F^{C_2}(2)F^{C_3}(3)F^{C_4}(4)>\; &=&
\;<F^{C_1}(1)F^{C_2}(2)>\;<F^{C_3}(3)F^{C_4}(4)>\; +\nonumber \\
&&\; <F^{C_1}(1)F^{C_3}(3)>\;<F^{C_2}(2)F^{C_4}(4)>\; +\nonumber \\
&&\;<F^{C_1}(1)F^{C_4}(4)>\;<F^{C_2}(2)F^{C_3}(3)>,\nonumber \\
{\rm where}\; F^{C_i}(i)\; \equiv \; F_{\mu_i\nu_i}^{C_i}(x_i,w)&&
\end{eqnarray}}\parindent0em
and analogously for all other expectation values. We obtain
{\mathindent-1em
\begin{eqnarray}
\label{Ber F 2}
\Delta F^2_{\alpha\beta}\:(x) = \frac{4}{R_P^4} \frac{1}{<W>} \Bigg(
\sum_{n=1}^\infty (-i)^{2n} \stackrel{\rm surface ordered}{\int \cdots \int}
d\sigma_{\mu_1\nu_1}^W\cdots d\sigma_{\mu_{2n}\nu_{2n}}^W\: tr\left[ {\bf
t}^{a_1}\cdots {\bf t}^{a_{2n}} \right]\times \nonumber \\
\int\int d\sigma_{\mu\nu}^P d\sigma_{\rho\sigma}^P
\frac{(-i)^2}{2!2} \Bigg[  \sum_{\rm pairs}^{\alpha_1,\beta_1\cdots
\alpha_n,\beta_n} <g^2 F_{\mu_{\alpha_1}\nu_{\alpha_1}}^{a_{\alpha_1}}
F_{\mu_{\beta_1}\nu_{\beta_1}}^{a_{\beta_1}}> \cdots\nonumber \\
<g^2 F_{\mu_{\alpha_{n-1}}\nu_{\alpha_{n-1}}}^{a_{\alpha_{n-1}}}
F_{\mu_{\beta_{n-1}}\nu_{\beta_{n-1}}}^{a_{\beta_{n-1}}}> \times <g^2
F_{\mu_{\alpha_n}\nu_{\alpha_n}}^{a_{\alpha_n}} F_{\mu\nu}^a>\; <g^2
F_{\mu_{\beta_n}\nu_{\beta_n}}^{a_{\beta_n}} F_{\rho\sigma}^a> \Bigg] \Bigg).
\label{Herl f 1}
\end{eqnarray}}\parindent0em
Of the $2n$ field strengths running over the pyramid of the loop $\cal C$ two
are correlated with a $F_{\mu\nu}^C$ on the pyramid of the plaquette and the
other $(2n-2)$ among one another. As soon as the two field strengths of one of
the latter $(n-1)$-pairs are not direct neighbors in surface ordering, this
ordering is suppressed by at least a factor $a/L$ as compared to the fully
ordered ones ($a$ is the correlation length and $L$ the linear extension of the
loop $\cal C$). We now neglect {\bf in any order of the expansion} of the
exponentials all terms which are not fully ordered, i.e.~which are suppressed
by at least a factor $a/L$. For all fully ordered expressions the traces in
eq.(\ref{Ber F 2}) are identical and one obtains:
{\mathindent0em
\begin{eqnarray}
\Delta F^2_{\alpha\beta}\:(x) = \frac{4}{R_P^4} \frac{1}{<W>} \Bigg(
\sum_{n=1}^\infty \frac{(-i)^{2n+2}\cdot N_c}{4(2N_c)^n(N_c^2-1)} \stackrel{\rm
surface ordered}{\int \cdots \int} d\sigma_{\mu_1\nu_1}^W\cdots
d\sigma_{\mu_{2n}\nu_{2n}}^W \times \nonumber \\
\int\int d\sigma_{\mu\nu}^P d\sigma_{\rho\sigma}^P  \Bigg[  \sum_{\rm
pairs}^{\alpha_1,\beta_1\cdots \alpha_n,\beta_n} <g^2
F_{\mu_{\alpha_1}\nu_{\alpha_1}}^{C_1} F_{\mu_{\beta_1}\nu_{\beta_1}}^{C_1}>
\cdots \\
<g^2 F_{\mu_{\alpha_{n-1}}\nu_{\alpha_{n-1}}}^{C_{n-1}}
F_{\mu_{\beta_{n-1}}\nu_{\beta_{n-1}}}^{C_{n-1}}> \times <g^2
F_{\mu_{\alpha_n}\nu_{\alpha_n}}^{C_n} F_{\mu\nu}^{C_n}>\; <g^2
F_{\mu_{\beta_n}\nu_{\beta_n}}^{C_{n+1}} F_{\rho\sigma}^{C_{n+1}}>\Bigg]
\Bigg)\nonumber
\end{eqnarray}}\parindent0em
where the $N_c$-factors are due to the traces.\\
Now the integrand is symmetric and we may replace the ordered surface
integrations by unconstrained ones, correcting for the larger phase space by
the factor $1/(2n)!$:
{\mathindent0em
\begin{eqnarray}
\Delta F^2_{\alpha\beta}\:(x)= \frac{4}{R_P^4} \frac{1}{<W>} \Bigg(
\sum_{n=1}^\infty \frac{(-i)^{2n+2}\cdot N_c}{4(2N_c)^n(N_c^2-1)(2n)!} \int
\cdots \int d\sigma_{\mu_1\nu_1}^W\cdots d\sigma_{\mu_{2n}\nu_{2n}}^W \times
\nonumber \\
\int\int d\sigma_{\mu\nu}^P d\sigma_{\rho\sigma}^P  \Bigg[  \sum_{\rm
pairs}^{\alpha_1,\beta_1\cdots \alpha_n,\beta_n} <g^2
F_{\mu_{\alpha_1}\nu_{\alpha_1}}^{C_1} F_{\mu_{\beta_1}\nu_{\beta_1}}^{C_1}>
\cdots \\
<g^2 F_{\mu_{\alpha_{n-1}}\nu_{\alpha_{n-1}}}^{C_{n-1}}
F_{\mu_{\beta_{n-1}}\nu_{\beta_{n-1}}}^{C_{n-1}}> \times <g^2
F_{\mu_{\alpha_n}\nu_{\alpha_n}}^{C_n} F_{\mu\nu}^{C_n}>\; <g^2
F_{\mu_{\beta_n}\nu_{\beta_n}}^{C_{n+1}} F_{\rho\sigma}^{C_{n+1}}> \Bigg]
\Bigg)\nonumber
\end{eqnarray}}\parindent0em
Performing the integration over each correlator separately we obtain:
{\mathindent-1em
\begin{eqnarray}
\Delta F^2_{\alpha\beta}\:(x) &=& \frac{4}{R_P^4} \frac{1}{<W>} \Bigg(
\sum_{n=1}^\infty \frac{(-i)^{2n+2}\cdot N_c}{4(2N_c)^n(N_c^2-1)2^{n-1}(n-1)!}
\times \nonumber \\
&&\left( \int\int d\sigma_{\mu\nu}^P d\sigma_{\rho\sigma}^W <g^2 F_{\mu\nu}^C
F_{\rho\sigma}^C > \right)^2\times \left( \int \int d\sigma_{\mu\nu}^W
d\sigma_{\rho\sigma}^W <g^2 F_{\mu\nu}^C F_{\rho\sigma}^C> \right)^{n-1} \Bigg)
\nonumber\\
&=& \frac{4(-i)^4}{8\cdot R_P^4(N_c^2-1)} \frac{1}{<W>} \left( \int\int
d\sigma_{\mu\nu}^P d\sigma_{\rho\sigma}^W <g^2 F_{\mu\nu}^C F_{\rho\sigma}^C >
\right)^2 \times \nonumber \\
&&\left(\sum_{n=0}^\infty \frac{(-i)^{2n}}{2^n n!(2N_c)^n}\left( \stackrel{\rm
frei}{\int \int} d\sigma_{\mu_1\nu_1}^W d\sigma_{\mu_{2}\nu_{2}}^W <g^2
F_{\mu\nu}^C F_{\rho\sigma}^C> \right)^n \right)
\end{eqnarray}}\parindent0em
In the approximation used the sum is just $\frac{<W>}{N_c}$ and the final
result is
\begin{equation}
\label{f Endergebnis}
\Delta F^2_{\alpha\beta}\:(x) \; = \; \frac{4}{R_P^4}
\frac{(-i)^4}{8N_C(N_C^2-1)} \left[ \int \int d\sigma_{\mu\nu}^W
d\sigma_{\rho\sigma}^P <g^2F_{\mu\nu}^C F_{\rho\sigma}^C > \right]^2\; .
\end{equation}
We note that in the integral one surface element is connected to the
Wegner-Wilson-loop (indicated by the index W) and one to the plaquette oriented
in $\alpha,\beta$-direction (indicated by P).

\epsfxsize5cm
\begin{figure}[ht]
\leavevmode
\centering
\epsffile{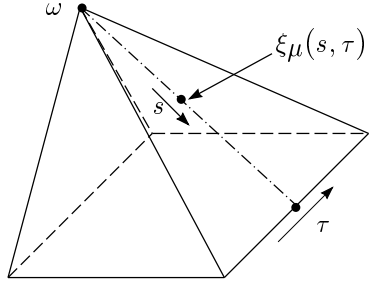}
\caption{Parametrization of the sliding sides of a pyramid. $s$ and $\tau \in
[0,1]$.}
\label{Parametrisierung}
\end{figure}\parindent0em
We now parametrize the surfaces as indicated in figure \ref{Parametrisierung}
and obtain with eq.(\ref{Korrelator}):
\pagebreak[3]
{\mathindent0em
\begin{eqnarray}
\Delta F^2_{\alpha\beta}\:(x) &=& \frac{4}{R_P^4} \frac{1}{(N_C^2-1)8N_C}
\Bigg[ \sum_{\rm surfaces} \int_0^1 ds_W \: ds_P \: d\tau_W \: d\tau_P
\frac{<g^2FF>}{12} \nonumber \\
&& \sum_{\mu < \nu \atop \rho < \sigma} \frac{\partial \left(
\xi^W_\mu(s_W,\tau_W), \xi^W_\nu(s_W,\tau_W) \right)}{\partial \left(
s_W,\tau_W \right) } \cdot \frac{\partial \left( \xi^P_\rho(s_P,\tau_P),
\xi^P_\sigma(s_P,\tau_P) \right)}{\partial \left( s_P,\tau_P \right) } \times
\nonumber \\
&& \bigg\{ \left( \delta_{\mu\rho} \delta_{\nu\sigma} -\delta_{\mu\sigma}
\delta_{\nu\rho} \right) \kappa D(z^2/a^2) + \\
&& \frac{1}{2}\left( \frac{\partial}{\partial z_\nu}(z_\sigma \delta_{\mu\rho}
- z_\rho \delta_{\mu\sigma})+\frac{\partial}{\partial z_\mu}(z_\rho
\delta_{\nu\sigma} - z_\sigma \delta_{\nu\rho}) \right) (1-\kappa )D_1(z^2/a^2)
\bigg\}
\Bigg]^2 \; \nonumber
\end{eqnarray}}\parindent0em
Here $z=\xi^W-\xi^P$ and the sum goes over the 16 different surface
combinations parametrized by $\xi^W_\mu(s_W,\tau_W)$ and
$\xi^P_\mu(s_P,\tau_P)$.\\
Since we are interested in the limit $R_P\to 0$ the integration over $\tau_P$
becomes trivial if we expand the expression around $\tau_P = 1/2$. We found
that in all cases the limit $R_P\to 0$ was practically reached for $R_P \leq
\frac{1}{6} a$.\\
Numerically the limit of static quarks was reached for $T \geq 6a$. But in
order to study the static case we took the limit $T\to \infty$ analytically.
\section{Results and Discussion}
\subsection{General results}
The reference point $w$ plays a crucial role in the non-Abelian Stokes-theorem
and it has to be common to both surfaces in order to apply the MSV. Its choice
influences the results. Even for the application of the MSV to calculate a
simple Wegner-Wilson-loop the resulting area law refers to the area including
the reference point \cite{31}. This is certainly one of the less pleasant
features of the model and a consequence of the approximation of the complicated
measure of the QCD-action by a Gaussian one. If we deform the surface,
necessarily higher cumulants must be occur, since the result has to be
independent of the surface. For the one loop case the MSV is in accordance with
phenomenology and lattice calculations, if the minimal surface is chosen. We
therefore have always chosen the minimal surface in the evaluation of a
Wegner-Wilson-loop by the MSV in the Gaussian form.\\
To calculate $\Delta F^2_{\alpha\beta}\:(x)$ we proceed the same way. We choose
the reference point in such a way, that the total resulting surface of the
sliding sides of the pyramids is minimal. Since the surface of the plaquette is
considered in the limit $R_P\to 0$, this means, that the reference point $w$
has to be in the plane of the loop $\cal C$. This will always be our standard
choice. In order to investigate the influence of the position , we will later
also study the case where the reference point has an arbitrary position on the
connection line between the plaquette $\cal P$ and the loop.\\
We first concentrate on the typical non-Abelian tensor structure determined by
the correlation function $D$.\\
It can be seen by symmetry arguments that for that case the difference of the
square of the magnetic field strengths, i.e.~the plaquette $P_{\alpha\beta}(x)$
with no time ($x_4$) component, vanishes identically. This means that the
magnetic background field is not affected by the static color charges.\\
The electric field perpendicular to the Wegner-Wilson-loop is also practically
not affected but only the difference of the squared electric field parallel to
the loop.\\
In both cases the squared electric field difference is negative, i.e.~the
presence of the static source diminishes the vacuum fluctuations.\\
In figure 8 we display the value of the squared field strength parallel to the
spatial loop extension $\Delta F^2_{34}\:(x)=-<g^2E_z^2(x_3,r)>_{\rm q
\bar{q}-vacuum}$ as a function of the coordinates $r= \sqrt{x_1^2+x_2^2}$ and
$x_3$ (the results are rotational invariant around the $x_3$-axes) for
different spatial separations $R_W$ (see figure \ref{Pyramide} for the choice
of coordinates).

The squared field strength reaches its saturation value $\Delta
F^2_{34}\:(x)\approx 14 {\rm \frac{GeV}{fm^3}}$ for a spatial extension of the
Wegner-Wilson-loop $R_W\approx 4a=1.4{\rm fm}$. The transversal extension of
the flux tube, defined by
\begin{equation}
r_{\rm MS}\equiv \sqrt{\frac{\int dr \: r \: r^2 \: \Delta
F^2_{34}\:(0,r)}{\int dr \: r \: \Delta F^2_{34}\:(0,r)}},
\end{equation}
is practically independent of $R_W$ and about $1.8$ times the correlation
length $a$ (see figure \ref{rMS}).

\setcounter{figure}{8}
\epsfxsize7cm
\begin{figure}[ht]
\leavevmode
\centering
\epsffile{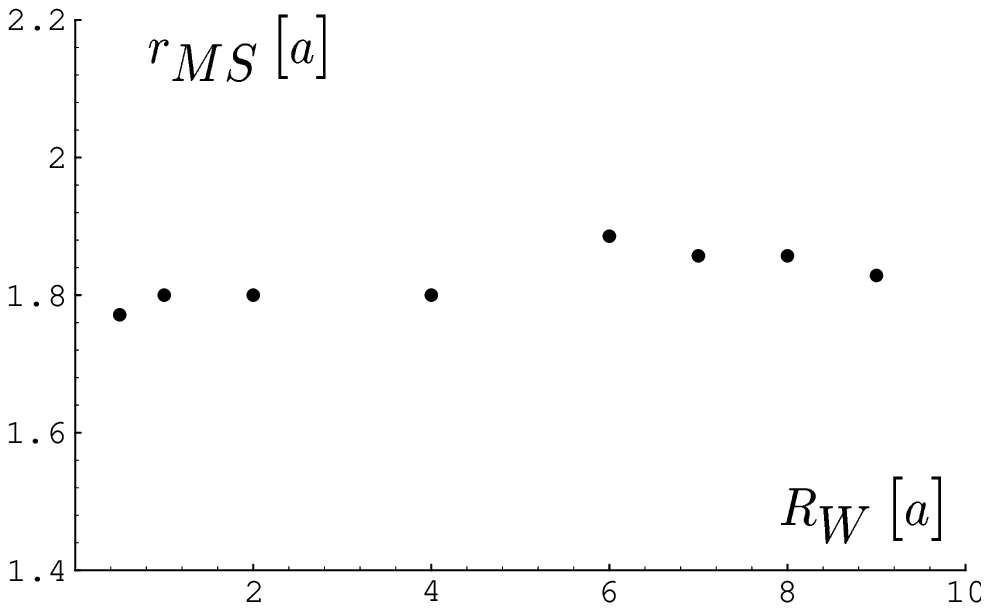}\\
\centering
\begin{minipage}{13cm}
\caption{The transversal extension $r_{\rm MS}$ of the squared electric field
as a function of the quark separation $R_W$.}
\label{rMS}
\end{minipage}
\end{figure}\parindent0em
As mentioned above, the quantity $\Delta F^2_{34}\:(x)$ is the difference
between the squared field strength in the presence of a static q-$\rm
\bar{q}$-pair and the non-perturbative vacuum field strength. In figure 10 we
display the unsubtracted squared field tensor of the color source in units of
the vacuum gluon condensate $<g^2FF>$.

As can be seen, the presence of the static source has only a moderate influence
on the global vacuum-structure ($\approx$10\%).\\
Now we discuss the second correlation function $D_1$. The contribution of this
correlator to $\Delta F^2_{\alpha\beta}\:(x)$ is numerically much less
important than that of the correlator $D$. For small distances this is mostly
due to the factor $(1-\kappa) = 0.26$, but for large distances the fact that
with this correlator no string is formed becomes decisive.\\
Again the square of the magnetic field is not influenced by the static source,
but now the squared parallel {\bf and} perpendicular electric fields are of the
same size and build up a spherical distribution. In figure 13 we show $\Delta
F^2_{14}\:(x_3,r)+\Delta F^2_{24}\:(x_3,r)+\Delta F^2_{34}\:(x_3,r)$ for the
correlator $D_1$.

\subsection{Michael sum rule}
The total field energy due to the presence of the static quark source is just
the potential energy of the quarks. We thus have a relation between the
expectation value of the Wegner-Wilson-loop and the spatial integral over the
sum of $\Delta F_{\alpha\beta}^2\:(x)$. This is one of the Michael sum rules
\cite{5}. In our case we can not test this sum rule absolutely, since we do not
know $g^2$. We can however estimate the effective value of $g^2$ from potential
models of quarkonia to $g^2 \approx 6$ \cite{13}. The Michael sum rule yields
than a very sensitive test for the some more technical approximations made to
calculate $\Delta F_{\alpha\beta}^2\:(x)$. We remind especially of the special
choice of factorization and the neglect of all terms suppressed by at least a
factor $a/L$ in the expansion of the Wegner-Wilson-loop.\\
In figure \ref{Egesamt} we display the numerically integrated energy density
$\frac{1}{2g^2}(\Delta F^2_{14}+\Delta F^2_{24}+\Delta F^2_{34})$ with
$g^2=7.2$ and compare it to the q-$\rm \bar{q}$-potential (\ref{Potential}) as
determined from the model of the stochastic vacuum.

\setcounter{figure}{13}
\epsfxsize7cm
\begin{figure}[ht]
\leavevmode
\centering
\epsffile{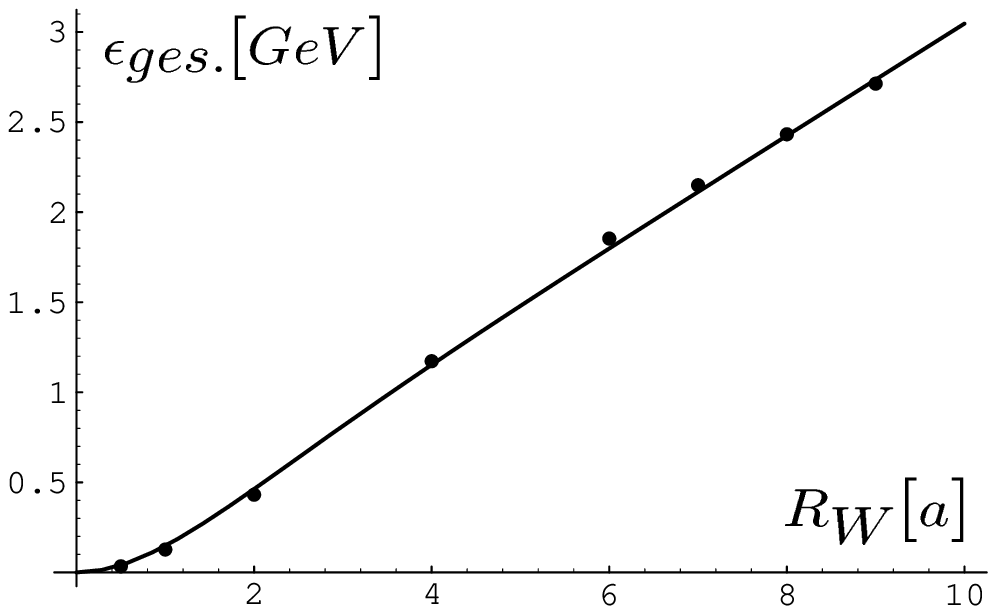}\\
\begin{minipage}{13cm}
\caption{The total energy stored in the field (dots) compared with the
potential of a q-$\rm \bar{q}$-pair as obtained by eq.(4)}
\label{Egesamt}
\end{minipage}
\end{figure}\parindent0em
We observe an almost perfect agreement between the two determinations even for
distances which are small as compared to the correlation length $a$. (The error
of the points is about the same size as the points due to the numerical
integration.) The agreement in form and the very reasonable value of
$\alpha_s\approx 0.57$ for the effective strong interaction coupling show, that
the approximations of the MSV are well consistent.
\subsection{Color-Coulomb contribution}
Having determined the effective $\alpha_s$ we are in the position to obtain an
approximation to the full QCD-expression by adding to the non-perturbative
results the color-Coulomb contribution. The result for the full energy density
is displayed in figure 17 for $R_W=2a$ and $R_W=9a$.

For $R_W=0.7$fm the perturbative contribution increases the energy density in
the middle of the flux tube from $0.65\rm \frac{GeV}{fm^3}$ to $4.2\rm
\frac{GeV}{fm^3}$ and even for $R_W=1.4$fm this increase goes from $1.1\rm
\frac{GeV}{fm^3}$ to $1.9\rm \frac{GeV}{fm^3}$. Since lattice calculations are
typically working with Wegner-Wilson-loops of spatial expansion $R_W\approx
1$fm the measured field strength is to a large extent a perturbative one. The
magnitude of our full energy densities (SU(3)) agrees qualitatively with the
scaled lattice results for SU(2) \cite{8} \cite{9} \cite{10}.
\subsection{Dependence of the results on the reference point}
In a non-Abelian gauge theory it is essential to transport the color content of
the field strength correlator to a fixed reference point in order to obtain
gauge covariant results. Such a reference point also occurs quite naturally in
the non-Abelian Stokes-theorem as starting point of the surface ordering
\cite{31}. The final result for the expectation value of a Wegner-Wilson-loop
must be independent of the choice of the reference point, but the result of the
MSV calculation does depend on it, if the reference point is moved outside the
surface under consideration. As discussed before the minimal surface is most
adequate for the simplification of the measure of the QCD-action to a Gaussian
one. Thus this has been our standard choice. Now we want to test the dependence
of the results on the reference point, varying its position on the connection
line between the plaquette and the loop as indicated in figure \ref{defalpha}.
$\alpha$=0 corresponds to our standard choice of $w!
$ on the loop and for $\alpha$=1 t
he reference point is inside the plaquette.

\setcounter{figure}{17}
\epsfxsize3cm
\begin{figure}[ht]
\leavevmode
\centering
\epsffile{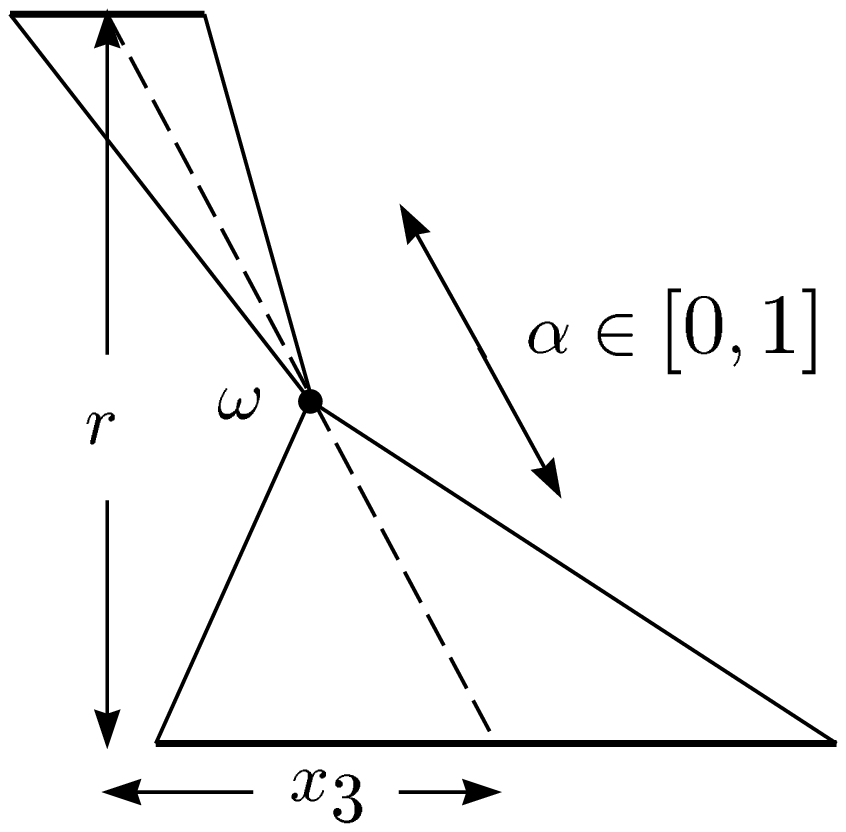}\\
\centering
\begin{minipage}{13cm}
\caption{Definition of the parameter $\alpha$ which describes the position of
the reference point $w$. $\alpha$=0 corresponds to our standard choice and for
$\alpha$=1 the reference point is inside the plaquette.}
\label{defalpha}
\end{minipage}
\end{figure}\parindent0em
In figure \ref{alpha6x10z0} we display the squared electric field parallel to
the quark axis in the transversal plane between the two quarks ($\Delta
F_{34}^2\:(0,r)$). In figure \ref{alpha6x10x1} we show the same field as a
function of the position between the quarks with a distance of one correlation
length from the q-$\rm \bar{q}$-axis ($\Delta F_{34}^2\:(x_3,1a)$). For the
parameter $\alpha$ we took the values
$\alpha=0,\frac{1}{4},\frac{1}{2},\frac{3}{4},1$.

\epsfxsize5.5cm
\begin{figure}[ht]
\leavevmode
\centering
\epsffile{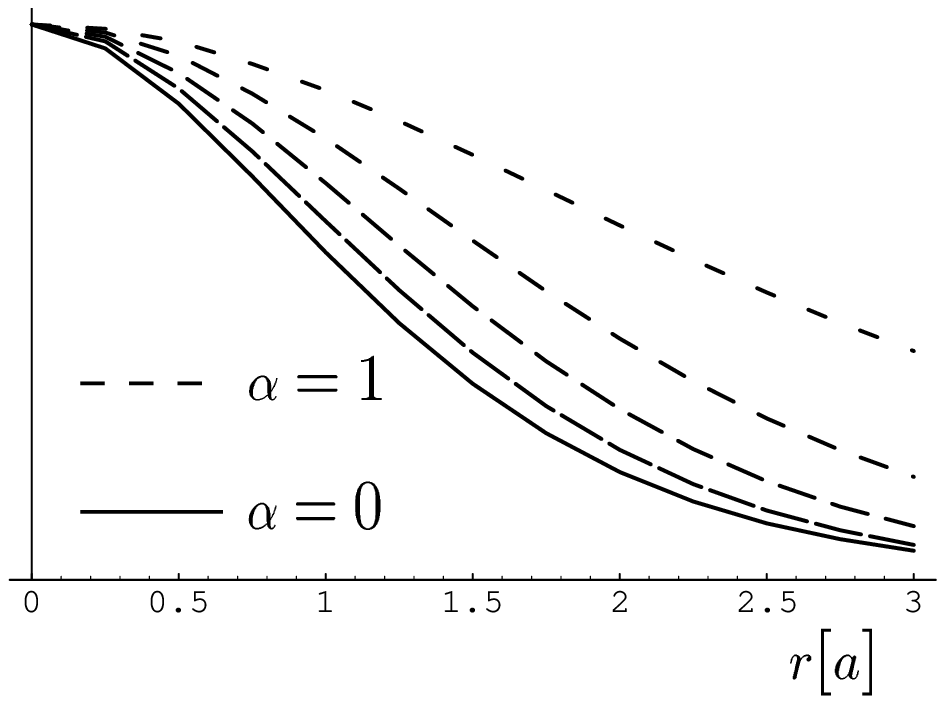}\\
\centering
\begin{minipage}{13cm}
\caption{The squared electric field parallel to the q-$\bar{\rm q}$-axis in the
transversal plane between the two quarks ($\Delta F_{34}^2\:(0,r)$) for a quark
separation $R_W=6a$ and for different positions of the reference point $w$ (see
figure 18). The solid line is for $\alpha = 0$ and the values $\alpha=
\frac{1}{4} ,\frac{1}{2},\frac{3}{4},1$ are plotted by subsequent shorter
dashes.}
\label{alpha6x10z0}
\end{minipage}
\end{figure}
\epsfxsize5.5cm
\begin{figure}[ht]
\leavevmode
\centering
\epsffile{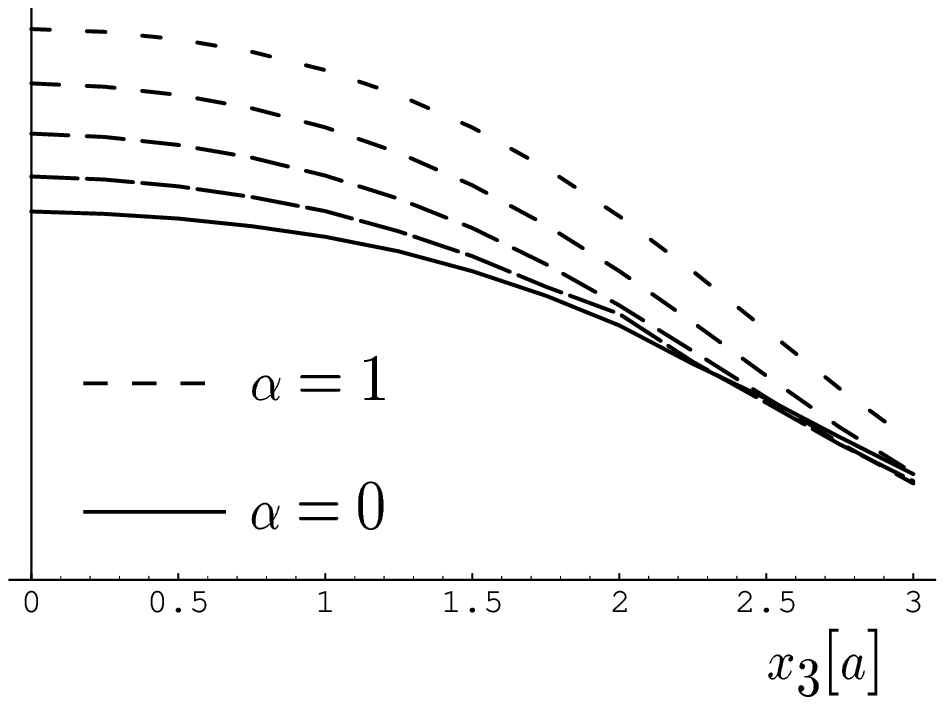}\\
\centering
\begin{minipage}{13cm}
\caption{The squared electric field parallel to the q-$\bar{\rm q}$-axis as a
function of the position between the quarks with a distance of one correlation
length from the q-$\rm \bar{q}$-axis ($\Delta F_{34}^2\:(x_3,1a)$) for a quark
separation $R_W=6a$ and for different positions of the reference point $w$ (see
figure 18).}
\label{alpha6x10x1}
\end{minipage}
\end{figure}\parindent0em
\newpage
As can be seen, the results are rather stable for $\alpha \le \frac{1}{2}$ and
the variation of $\alpha$ has only a limited numerical effect.
\subsection{Summary}
The MSV in its simplest form makes definite predictions on the non-perturbative
contributions of the chromodynamic fields of a static q-$\rm \bar{q}$-pair. The
first result is that the expectation values of the squared magnetic and
perpendicular electric field are not affected by the quark pair whereas the
squared electric field parallel to the quark axis is diminished. This is in
qualitative agreement with lattice gauge calculations of the connected linear
field strengths \cite{2}. The vacuum structure is quite stable, because the
fluctuations of $<g^2FF>$ are decreased by only 10\%.\\
With increasing quark separation a flux tube with constant width ($r_{\rm
MS}\approx 0.64$fm) and energy density of about $1 \frac{\rm GeV}{\rm fm^3}$ is
built.\\
The approximations in the MSV made here respect the Michael sum rule with an
effective strong coupling $\alpha_S$=0.57. By adding the perturbative
contribution with this coupling to the non-perturbative one we obtained the
total energy density in agreement with lattice SU(2) gauge calculations
\cite{8} \cite{9} \cite{10}.

{\large \bf Acknowledgements}

We would like to thank O. Nachtmann for fruitful discussions.

\end{document}